\documentclass[review]{elsarticle}

\usepackage{lineno,hyperref}
\modulolinenumbers[1]

\journal{arXiv}

\begin{document}

\begin{frontmatter}

\title{Estimation of Radio Refractivity from a Decade Satellite Derived Meteorological Data for West Africa}

\author[mymainaddress]{O. J. Abimbola\corref{mycorrespondingauthor}}
\cortext[mycorrespondingauthor]{Corresponding author}
\ead{abimbola.oladiran@science.fulafia.edu.ng}

\author[mymainaddress]{E. Bada}
\author[mysecondaryaddress]{A. O. Falaiye}
\author[mysecondaryaddress]{Y. M. Sukam}
\author[mymainaddress]{S. Muhammad}

\address[mymainaddress]{Department of Physics, Federal University of Lafia, Nigeria}
\address[mysecondaryaddress]{Department of Physics, University of Ilorin, Nigeria}

\begin{abstract}
Radio refractivity, which is the bending of radio signal as it propagates through media, is very important in works involving terrestrial atmospheric electromagnetic propagation such as point-to-point microwave communication, terrestrial radio and television broadcast, mobile communication system, and so on. This study has focused on the West African region where it was found that the region has a refractivity which varies exponentially with height and from the coast towards the desert. Generally, refractivity gradient was found to range between -46.48 and -29.51 \textit{N-units/km} (\textit{k}-factor value of between  1.23 and 1.42) across the region, splitting the between sub- and super-refraction. The variation in refraction type was found to follow seasonal pattern across the West African region, with sub-refraction dominating during dry season and super-refraction dominating most part in the coastal area during the wet season.
\end{abstract}

\begin{keyword}
radio refractivity\sep West Africa\sep refractivity gradient\sep \textit{k}-factor
\end{keyword}

\end{frontmatter}


\section{Introduction}

The behavior of radio waves in the tropospheric layer of the Earth’s atmosphere is very important in this modern age that is highly influenced by radio communications ranging from mobile telephoning through terrestrial digital broadcasting to the propagation of satellite radio signal through the troposphere.

Propagation of radio waves in the inhomogeneous troposphere leads to the refraction of the radio waves with subsequent refractive index \textit{n}, given by
\begin{equation}
	n=\frac{speed ~of ~EM-waves ~in ~air}{speed ~of ~EM-waves ~in ~vacuum}
	\label{eq1}
\end{equation}
The refractive index \textit{n}, usually have a very small value for the troposphere (1.0000 to 1.0000004445). In order to have an easy to work with value, \textit{n} is usually transformed to the refractivity \textit{N} by \cite{Ko}
\begin{equation}
	N=\left(n-1\right)\times10^{6}
	\label{eq2}
\end{equation}
Since n is a unit-less quantity, \textit{N} is also expected to be unit-less, nevertheless, \textit{N} is usually given in the unit of “\textit{N-unit}”. In the troposphere, \textit{N} ranges between 200 to 400 \textit{N-units}.

According to the International Telecommunication Union recommendation \cite{ITU-R}, the radio refractivity index \textit{N}, can be given as \textit{N(p, e, T),} where \textit{p} is the total atmospheric pressure, \textit{e} is the vapour pressure and \textit{T} is the ambient temperature:
\begin{equation}
	N=77.6\frac{p}{T}+3.75\times10^{5}\frac{e}{T^{2}}
	\label{eq3}
\end{equation}
where, \textit{p} and \textit{e} are measured in \textit{hPa} and \textit{T} is measured in \textit{Kelvin}. As is obvious from Equation \ref{eq3}, atmospheric radio refractivity depends significantly on the atmospheric moisture content. It should be noted that Equation \ref{eq3} is valid for radio frequencies up to $100 ~GHz$ \cite{ITU-R, Babin, Adediji}.

In grouping the radio refractivity into groups such as normal/standard refraction, sub-refraction, super-refraction and ducting, a quantity known as the effective earth radius factor or \textit{k}-factor, is employed. The \textit{k}-factor is given as
\begin{equation}
	k=\frac{1}{1+\left(dN/dh\right)/157}
	\label{eq4}
\end{equation}
here, $dN⁄dh=G$ which is known as the refractivity gradient \cite{Adediji, Hall, Afullo}. \textit{G} is a quantity which is very useful in describing radio refractivity at the first $1 ~km$ of the troposphere, from the earth’s surface \cite{ITU-R}: if $G > -40$ then refraction is sub-refraction; if $G < -40$ then refraction is super-refraction while if $G < -157$ the refraction is termed ducting \cite{AdedijiA}. In terms of earth radius factor, standard refractivity occurs at $k = 1.33$; sub-refraction, that is, bending of the radio waves away from the earth’s surface, occurs at $1.33 < k < \infty$; bending of the radio waves towards the earth, known as super-refraction, occurs at $0 < k < 1.33$ while, ducting occurs at $-\infty < k < 0$  \cite{Adediji, Aremu}.

Several investigators have worked on the study of radio refractivity using localized data from radiosonde and masts. Davis et al. \cite{Davis} used ground-based microwave radiometer to study the local spatio-temporal variation of the radio refractivity in west coast of Sweden, they found an average surface wet-refractivity horizontal gradients of between 0.1 – 1.0 $N-units/km$ by assuming exponential profile for the wet-refractivity horizontal gradient, however they noted that their model will be inadequate for period of time longer than 30 $min$.  Focusing on some locations within the North Central Nigeria, Emetere \textit{et al}. \cite{Emetere} studied the effects of aerosol on the radio refractivity. Ojo \textit{et al}. \cite{Ojo}, studied geoclimatic variation of radio refractivity across Nigeria where map of \textit{k}-factor for Nigeria was presented and they generally found that \textit{k}-factor always exceed the value of 1.33 in Nigeria.

It is our aim, in this study, to use a ten-year satellite derived data to describe both horizontal and vertical distribution of radio refractivity in West Africa together with the description of refractivity gradient and the earth radius factor.

\section{Methodology}
West African region under consideration is between $18 ~^{o}W$ and $18 ~^{o}E$ longitudes. This region is a low latitude region between latitude $3 ~^{o}N$ and $18 ~^{o}N$. The study area generally has two distinct seasons: the wet season and the dry season. The wet season is a period of high atmospheric humidity and rainfall, the period vary across the region, at the coast the wet season could occur within the months of April to October while at places closer to the Sahara Desert the wet season could be from June to September. The remaining part of the year is marked by low atmospheric humidity and practically no rainfall, the early part of the dry season, within November, December and January, there is heavy dust lifting from the Sahara Desert and blown across the region, this dust movement period is called Harmattan period: this seasonal variation of atmospheric moisture significantly affects radio refractivity.
\begin{figure*}[!h]
	\centering
		\includegraphics[width=1.00\textwidth]{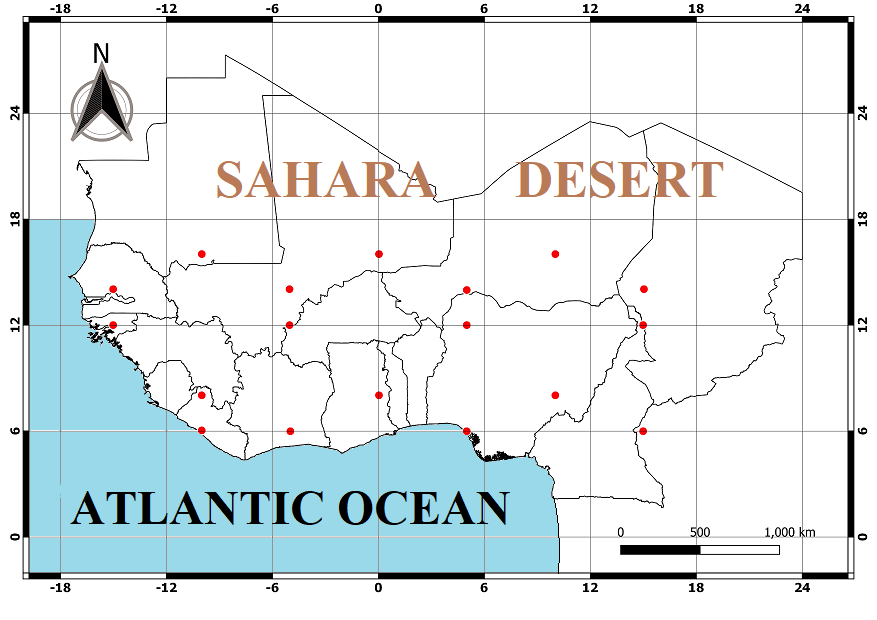}
	\caption{Map of West Africa showing the stations used for this study, marked in red dots}
	\label{fig:fig1}
\end{figure*}

Ambient temperature and specific humidity data were obtained from the archive of the Climate Monitoring Satellite Application Facility (CM-SAF) of the European Center for Medium Range Weather Forecast (ECMWF) at https://wui.cmsaf.eu/safira/action while the latitudinal radio refractivity was obtained from the Radio Occultation Meteorology Satellite Application Facility (ROM SAF) at https://romsaf.org. CM-SAF data was obtained at six different vertical pressure levels: 200 \textit{hPa}, 300 \textit{hPa}, 500 \textit{hPa}, 700 \textit{hPa}, 850 \textit{hPa} and 1000 \textit{hPa}, as a daily average for the years 2007 to 2016. The data from CM-SAF was obtained in netCDF format and python script together with Microsoft Excel software was used to extract the data while Equation \ref{eq5} was used to calculate the water vapour partial pressure \textit{e}
\begin{equation}
	e=\frac{0.001pq}{0.622}
	\label{eq5}
\end{equation}
where, \textit{p} = total atmospheric pressure; \textit{q} = specific humidity. All geo-plots were done using QGIS freeware.

\section{Results and Discussion}
Figure 2 shows the spatial distribution of the average radio refractivity for the years 2006 to 2016 at four different pressure levels. Figure 3 shows the latitudinal vertical variation of radio refractivity during the year 2018 for the months of January, April, August and December: generally in West Africa, December and January are within the dry very low humidity period of the year while April is the beginning of the wet humid season and August is about the peak of the wet humid season. Exponential vertical variation of radio refractivity at latitudes $2.5 ~^{o}N, 7.5 ~^{o}N, 12.5 ~^{o}N, 17.5 ~^{o}N$ and $22.5 ~^{o}N$, is clearly visible on Figure 4. 
\begin{figure*}[!h]
	\centering
		\includegraphics[width=1.00\textwidth]{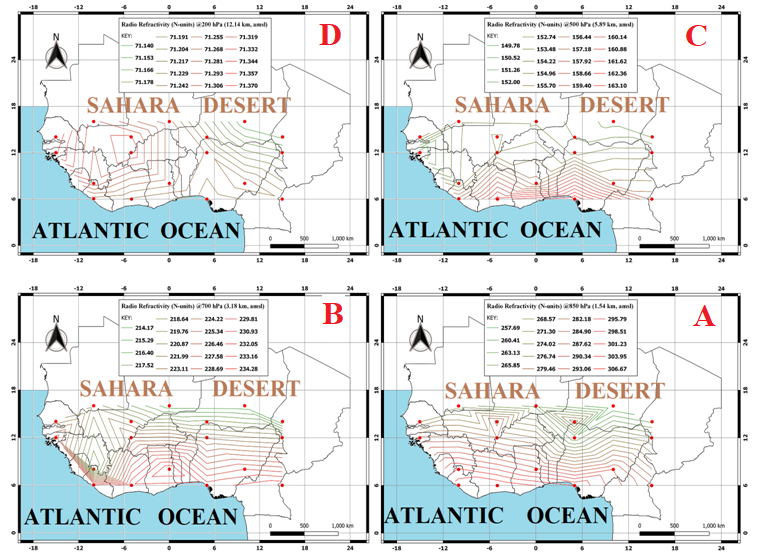}
	\caption{Spatial distribution of radio refractivity at different pressure level within the West African region: (A) at 850 \textit{hPa} pressure level (that is, around 1.54 \textit{km} amsl), (B) at 700 \textit{hPa} pressure level (that is, around 3.18 \textit{km} amsl), (C) at 500 \textit{hPa} pressure level (that is, around 9.89 \textit{km} amsl), (D) at 200 \textit{hPa} pressure level (that is, around 12.14 \textit{km} amsl).}
	\label{fig:fig2}
\end{figure*}

At 850 \textit{hPa}, that is about 1.54 \textit{km} above mean sea level, as presented on Figure 2A, the radio refractivity was found to range between 306.67 and 257.69 \textit{N-units} with a range of 48.98 \textit{N-units}, the radio refractivity could be observed to be higher at the coast and gradually decreases as we move to a dryer locations closer to the Sahara Desert, this is as a result of the dependence of radio refractivity on the atmospheric moisture burden. This variation in radio refractivity is replicated across other pressure levels but at 200 \textit{hPa}, that is, about 12.14 \textit{km} above mean sea level, where the radio refractivity ranges between 71.37 and 71.14 \textit{N-units} with a range of 0.23 \textit{N-units}, the radio refractivity was found to be slightly higher east of the prime meridian than to the west of the prime meridian: the coastal bound at the western part is more than at the eastern part of the West African region resulting in slightly more moisture at this altitude on the western part than at the eastern part.
\begin{figure*}[!h]
	\centering
		\includegraphics[width=1.00\textwidth]{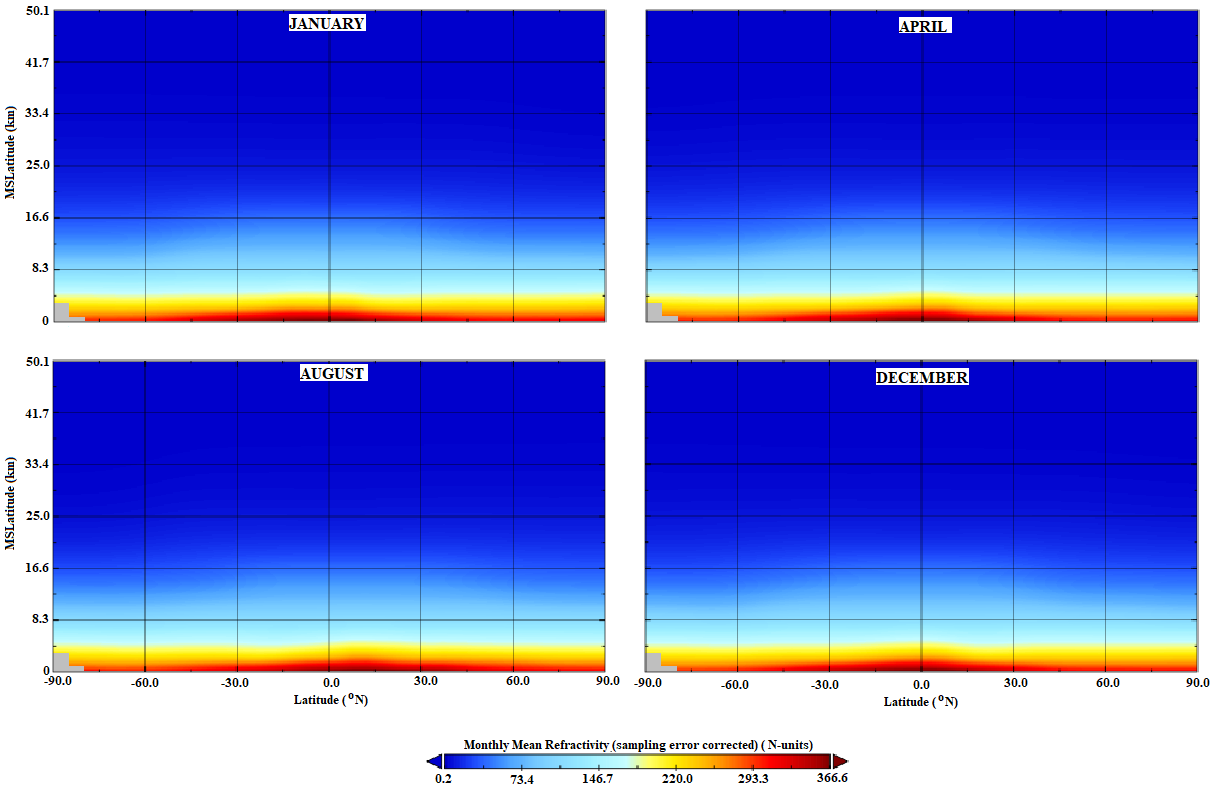}
	\caption{Vertical distribution of mean monthly radio refractivity, along the latitudes for the months of January, April, August and December in the year 2018.}
	\label{fig:fig3}
\end{figure*}

The vertical variation of the radio refractivity along the latitudes that includes the West Africa region (i.e., 2.5, 7.5, 12.5, 17.5 and 22.5 $^{o}N$) is as shown on Figure 3 and the mathematical models for the vertical variations is as shown on Table 1. From Figure 4, it could be observed that the vertical variations are exponential in nature which is what is obtained and shown on Table 1. The surface value of radio refractivity was found to range between 403 and 420 \textit{N-units} while the radio refractivity scale height was found to be around 6.6 \textit{km}. It should be noted that the models shown on Table 1 are for the latitudes and not regional, besides the table uses ROM-SAF data.
\begin{figure*}[!h]
	\centering
		\includegraphics[width=1.00\textwidth]{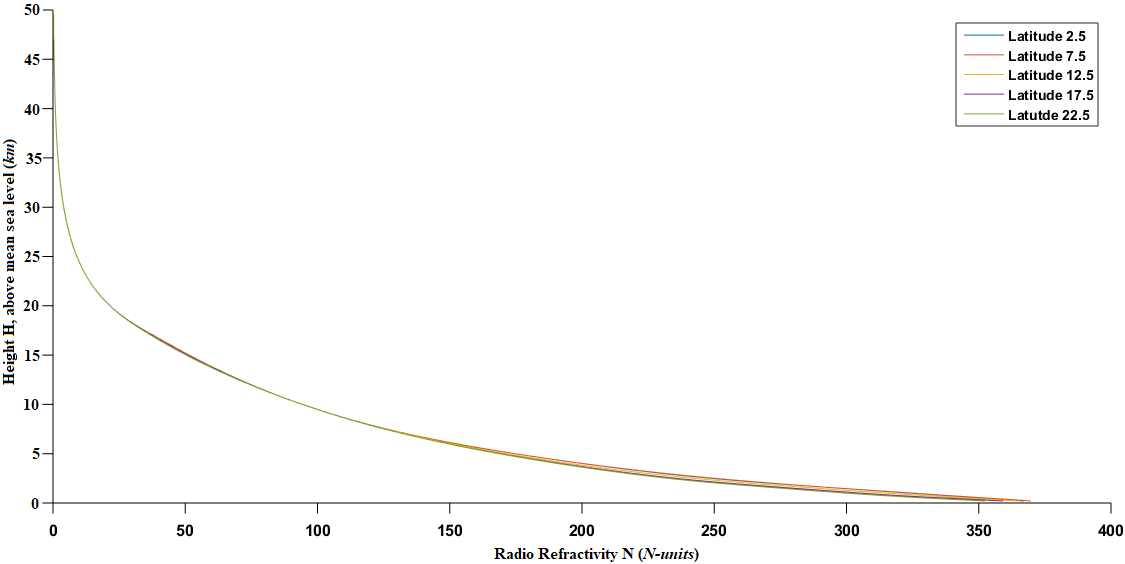}
	\caption{Vertical variation of the radio refractivity at several latitudes (ROM-SAF data).}
	\label{fig:fig4}
\end{figure*} 

\begin{table}[!t]
\centering
\caption{Mathematical models describing the vertical variation of the radio refractivity at different latitudes (ROM-SAF data).}
\label{tab1}
\begin{tabular}{lcc}   \hline
\bf{Latitude ($^{o}N$)} &	\textbf{Model}	& \textbf{Coefficient of} \\
	&		&\textbf{Determination ($R^{2}$)}\\ \hline
2.5   &	$N=420e^{-\left(H/6.6\right)}$	& 0.998\\
7.5  &	$N=420e^{-\left(H/6.6\right)}$	& 0.998\\
12.5  &	$N=415e^{-\left(H/6.6\right)}$	& 0.998\\
17.5   &	$N=415e^{-\left(H/6.6\right)}$	&	0.997	\\
22.5  &	$N=403e^{-\left(H/6.6\right)}$	&	0.998	\\
\hline
\end{tabular}
\end{table}

\begin{table}[!t]
\centering
\caption{Mathematical models describing the vertical variation of the radio refractivity at different locations within the West African region (CM-SAF data).}
\label{tab2}
\begin{tabular}{lccc}   \hline
\textbf{Latitude}&\bf{Longitude} &	\textbf{Model}	& \textbf{$R^{2}$} \\ \hline
16$~^{o}N$ &10$~^{o}E$    &	$N=313e^{-\left(H/8.33\right)}$	& 0.999\\ 
6$~^{o}N$ &5$~^{o}E$   &	$N=363e^{-\left(H/7.69\right)}$	& 0.999\\
6$~^{o}N$ &5$~^{o}W$   &	$N=367e^{-\left(H/7.69\right)}$	& 0.998\\
16$~^{o}N$ &10$~^{o}W$    &	$N=326e^{-\left(H/8.33\right)}$	&	0.998	\\
\hline
\end{tabular}
\end{table}

Using  CM-SAF data, location based vertical models for the radio refractivity were obtained for locations West and East of the prime meridian within the West African region, as presented in Table 2. Within the considered locations, the radio refractivity scale height was found to range between 7.28 and 8.19 \textit{km}, while the surface refractivity ranges between 315 and 368 \textit{N-units}, it is obvious from both Table 1 and 2 that the vertical variations is exponential in nature.

\subsection{Refractivity Gradient and k-factor}
A ten year average of the refractivity gradient across West Africa for the first 1.5 \textit{km} of the atmosphere, from the surface, is shown in Figure 5. From Figure 5, it could be seen that two major types of refraction dominates the atmosphere of West Africa, namely sub-refraction ($G > -40 ~N-units/km ~or ~\infty < k < 1.33$) and super-refraction ($G < -40 ~N-units/km ~or ~0 > k > 1.33$). Figure 5B shows finer details than Figure 5A, it could be observed that Figure 5B shows more area with super-refraction.

Taking some countries as point of focus, for instance Nigeria, we could see, on both Figure5A and B that the country is roughly divided into two parts, the northern half is dominated by sub-refraction while the southern part is dominated by super-refraction; this is in agreement with the work of Aremu et al. \cite{Aremu}, who worked in Ibadan, a southern Nigerian city and found \textit{G} = -50.42 \textit{N-units/km} as well as \textit{k} = 1.53, also Adediji et al. \cite{Adediji} working in Akure, the same southern part of Nigeria, found a two year average values of \textit{G} and \textit{k} to be -52.8 \textit{N-units/km} and 1.51 respectively. Fashade et al. \cite{Fashade}, using a five year satellite data found that Sokoto and Maiduguri, two cities in northern Nigeria has average \textit{G} values of -26.4 and -23.2 \textit{N-units/km} respectively while Abuja, Enugu, Lagos and Port Harcourt, cities in the middle belt (Abuja) and southern Nigeria were found to have average G values of -40.2, -40.3, -47.0 and -43.0 \textit{N-units/km} respectively. This result of Fashade et al. \cite{Fashade} agrees very well with the result presented on Figure 5. 
\begin{figure*}[!h]
	\centering
		\includegraphics[width=1.00\textwidth]{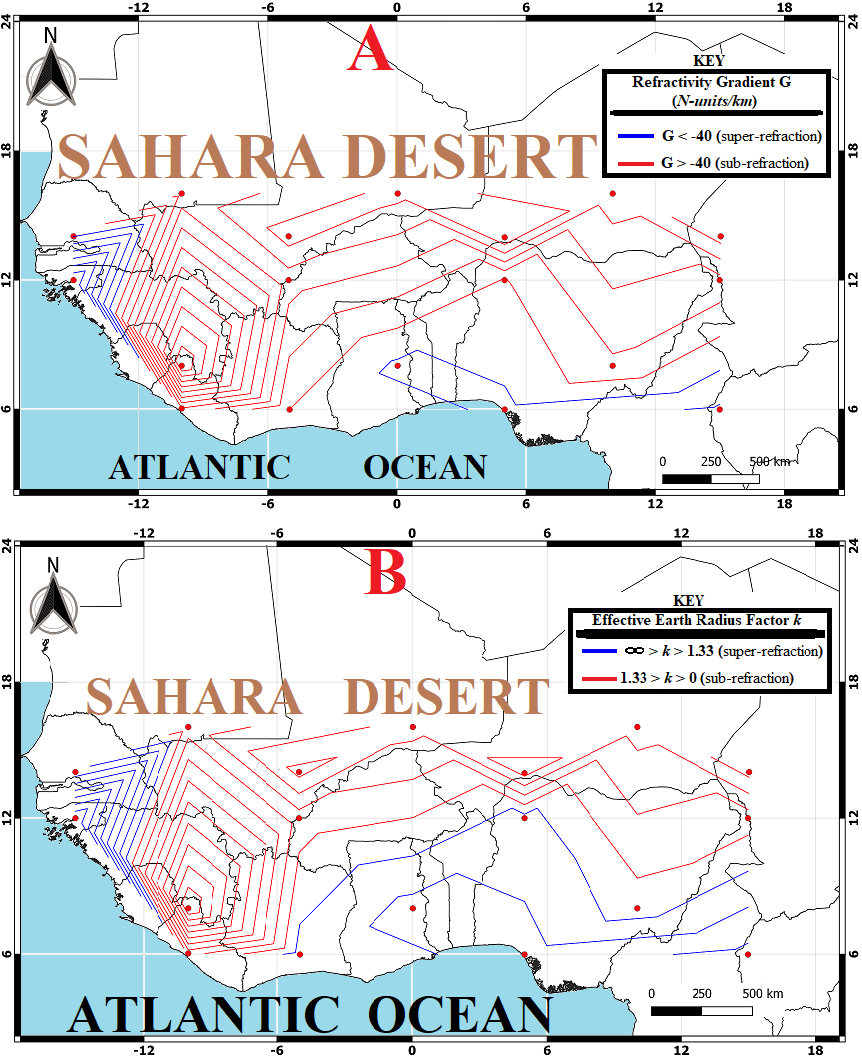}
	\caption{(A) West African refractivity gradient \textit{G} map and (B) earth radius factor \textit{k}  map, for the first 1.5 \textit{km} of the atmosphere.}
	\label{fig:fig5}
\end{figure*}

\begin{figure*}[!h]
	\centering
		\includegraphics[width=1.00\textwidth]{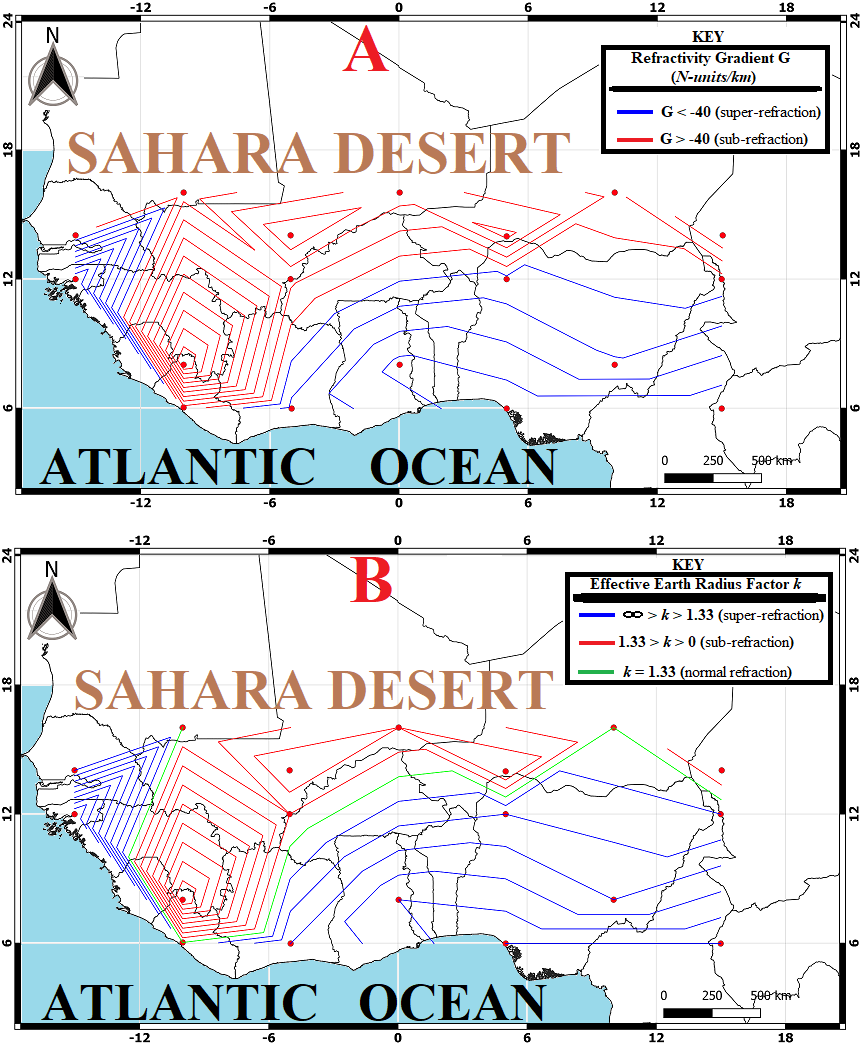}
	\caption{Wet Season map of (A) West African refractivity gradient \textit{G} map and (B) earth radius factor \textit{k} map, for the first 1.5 \textit{km} of the atmosphere.}
	\label{fig:fig6}
\end{figure*}

If Figure 5B is used for study, Senegal, Gambia, Guinea-Bissau and the western part of Guinea could be observed to be generally dominated by super-refraction, this is also true of Benin, Togo, Ghana and eastern part of Cote de Voire. The remaining countries of West Africa are then found to be dominated by sub-refraction, the region with sub-refraction have values of \textit{k}-factor ranging between 1.23 and 1.32 (\textit{G} ranges between -39.88 and -29.51 \textit{N-units/km}) while the region of super-refraction have \textit{k}-factor ranging between 1.34 and 1.42 (\textit{G} ranges between -46.48 and -40.82 \textit{N-units/km}). The regions of sub- and super-refractions could be attributed to the atmospheric moisture distribution in the West Africa region; the region of super-refraction could be observed to be close to the Atlantic Ocean and hence have higher atmospheric moisture burden while the region dominated by sub-refraction is a region much closer to the Sahara Desert, hence with drier atmosphere.

\subsection{Seasonal Variations of Refractivity Gradient and k-factor}
Figure 6 shows the distribution of refractivity gradient and \textit{k}-factor across West Africa during the wet season months of May, June, July, August, September and October. As could be observed on Figure 6B, the whole of Nigeria, Benin, Togo and Ghana as well as major part of Cote de Voire, Senegal, Gambia, and Guinea-Bissau are having super-refraction while all other part of the region is dominated by sub-refraction. The area of sub-refraction was found to have \textit{k}-factor ranging between 1.24 and 1.32 (\textit{G} value between -39.41 and -30.22 \textit{N-units/km}) while the area of super-refractivity have \textit{k}-factor of between 1.34 and 1.42 (\textit{G} value of between -46.76 and -40.33 \textit{N-units/km}). A very thin region of normal refractivity (\textit{k} = 1.33, \textit{G} = -40 \textit{N-units/km}) could be observed on Figure 6B passing through southern Niger, part of north-western Nigeria, Burkina Faso, Cote de Voire, Liberia, Sierra Leone, Guinea and south-western Mali.

\begin{figure*}[!h]
	\centering
		\includegraphics[width=1.00\textwidth]{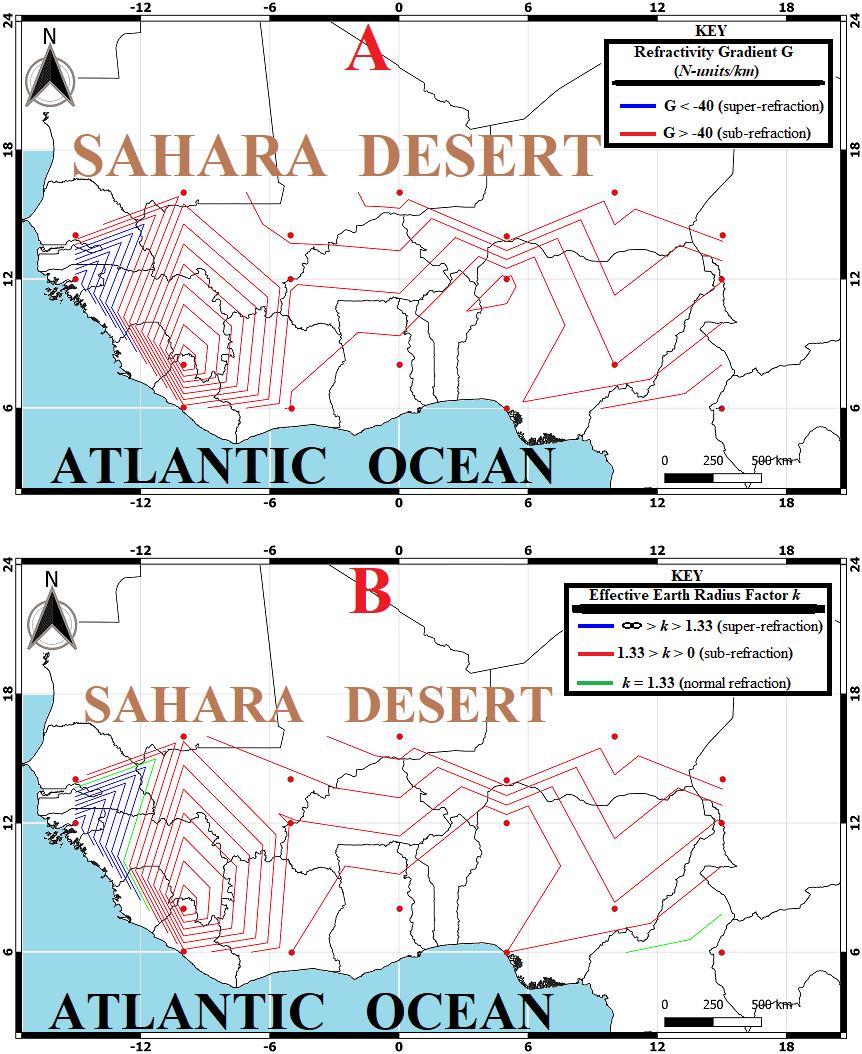}
	\caption{Dry Season map of (A) West African refractivity gradient \textit{G} map and (B) earth radius factor \textit{k} map, for the first 1.5 \textit{km} of the atmosphere.}
	\label{fig:fig7}
\end{figure*}

By the dry season months of November, December, January, February, March and April, the West African region is almost dominated by sub-refraction with only small region of Senegal, Gambia, Guinea Bissau and Guinea showing some level of super-refraction, there is also a very tiny region of normal refraction as shown on Figure 7B, during the West African dry season the atmosphere usually contains very little amount of water vapour. The region dominated with sub-refraction was found to have \textit{k}-factor within 1.22 and 1.32 (\textit{G} value of between -39.09 and -27.93 \textit{N-units/km}) while the region of super-refraction was found to have \textit{k}-factor of between 1.34 and 1.42 (\textit{G} value of between -46.20 and -40.11 \textit{N-units/km}).
\begin{figure*}[!h]
	\centering
		\includegraphics[width=1.00\textwidth]{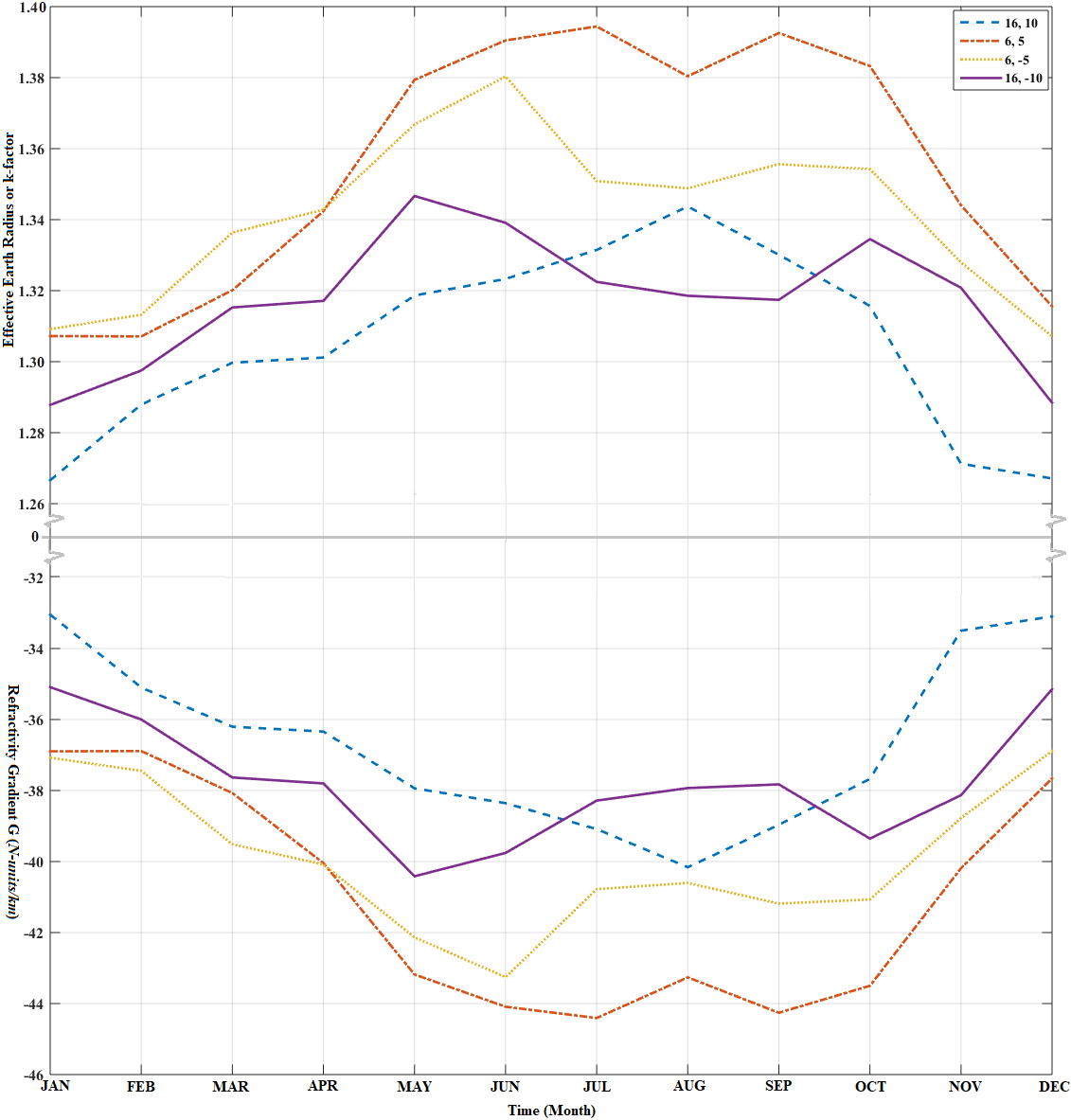}
	\caption{Time series plot of refractivity gradient \textit{G} and the \textit{k}-factor for four locations ($16 ~^{o}N, 10 ~^{o}W; 6 ~^{o}N, 5 ~^{o}W;  6 ~^{o}N, 5 ~^{o}E; 16 ~^{o}N, 10 ~^{o}E$)  within the West African region.}
	\label{fig:fig8}
\end{figure*}

A ten-year monthly average of \textit{k}-factor and refractivity gradient \textit{G} for four locations in West Africa region is presented on Figure 8. The four locations are chosen such that two are on either sides of the prime meridian ($5 ~^{o}E, 10 ~^{o}E$ and $5 ~^{o}W, 10 ~^{o}W$) as well as two at the lower latitude and two at the higher latitude ($6 ~^{o}N, 16 ~^{o}N$). It could be observed on Figure 8 that the \textit{k}-factor rises through the wet season while the refractivity gradient falls through the wet season, following variation in moisture through the seasons. Figure 8 confirms the fact that the lower latitudes in West Africa generally have super-refraction while the higher latitudes have sub-refraction.

\section{Conclusion}
Study of the radio refractivity, which is the change in the path of radio signal as it passes through a medium, is an important part of modern technology as it enhances modern communication, especially close to the earth’s surface such as point-to-point terrestrial microwave communication, terrestrial radio and television communication, wifi signals, and so on, also important in the study of satellite signal propagation through the atmosphere.

The vertical variation of radio refractivity in the region was found to follow an exponential model, that is, refractivity decreases exponentially with height. The refractivity was also found to be higher at locations closer to the ocean than locations closer to the desert. The current study has focused on West African region within the latitudes $4 ~^{o}N$ and $18 ~^{o}N$, as well as longitudes $18 ~^{o}W$ and $18 ~^{o}E$. The lower latitudes were found to be dominated by super-refraction while the higher latitudes were dominated by sub-refraction. The refractivity variation was found to be in synchronization with the seasonal variation in atmospheric moisture across the region.

\section*{Acknowledgements}
The authors wish to acknowledge the Radio Occultation Meteorology Satellite Application Facility (ROM SAF) and Climate Monitoring Satellite Application Facility (CM-SAF) of the EUMETSAT (data doi.org/10.5676/EUM\_SAF\\ \_CM/WVT\_ATOVS/V001) for making their data freely available for use. 


\begin{thebibliography}{10}
\expandafter\ifx\csname url\endcsname\relax
  \def\url#1{\texttt{#1}}\fi
\expandafter\ifx\csname urlprefix\endcsname\relax\def\urlprefix{URL }\fi
\expandafter\ifx\csname href\endcsname\relax
  \def\href#1#2{#2} \def\path#1{#1}\fi

\bibitem{Ko}
H.~W. Ko, J.~W. {Sari}, J.~P. {Skura}, Anomalous microwave propagation through
  atmospheric ducts, Johns Hopkins APL Technical Digest 4~(1) (1993) 12--26.

\bibitem{ITU-R}
ITU-R, The radio refractive index: its formula and refractivity data. itu radio
  communication assembly, ITU-R P—Series~(453-11).

\bibitem{Babin}
S.~M. Babin, G.~S. {Young}, J.~A. {Carton}, A new model of the oceanic
  evaporation duct, Journal of Applied Meteorology 36 (1997) 193--204.

\bibitem{Adediji}
A.~T. Adediji, M.~O. {Ajewole}, S.~E. {Falodun}, Distribution of radio
  refractivity gradient and effective earth radius factor (k-factor) over
  akure, south western nigeria, Journal of Atmospheric and Solar-Terrestrial
  Physics 73 (2011) 12--26.
\newblock \href {http://dx.doi.org/10.1016/j.jastp.2011.06.017}
  {\path{doi:10.1016/j.jastp.2011.06.017}}.

\bibitem{Hall}
M.~P.~M. Hall, Effect of the troposphere on radio communication, ieee
  electromagnetic wave series, Peter Peregrinus Ltd. United Kingdom (1989)
  105--116.

\bibitem{Afullo}
T.~J. Afullo, T.~{Motsoela}, D.~F. {Molotsi}, Refractivity gradient and
  k-factor in botswana, Radio Africa (1999) 107--110.

\bibitem{AdedijiA}
A.~T. Adediji, M.~O. {Ajewole}, Vertical profile of radio refractivity gradient
  in akure south-west nigeria, Progress in Electromagnetics Research C 4 (2008)
  157--168.

\bibitem{Aremu}
O.~A. Aremu, L.~O.~A. {Oyinkanola}, A.~{Akande}, W.~A. {Azeez}, Effects of
  radio refractivity gradient and k-factor on radio signal over ibadan, south
  western, nigeria, Global Scientific Journal 6~(5) (2018) 248--253.

\bibitem{Davis}
J.~L. Davis, G.~{Elgered}, A.~E. {Niell}, C.~E. {Kuehn}, Ground-based
  measurement of gradients in the wet radio refractivity of air, Radio Science
  28~(6) (1993) 1003--1018.
\newblock \href {http://dx.doi.org/0048-6604/93/93RS-01917508.00}
  {\path{doi:0048-6604/93/93RS-01917508.00}}.

\bibitem{Emetere}
M.~E. Emetere, O.~A. {Akinwumi}, T.~V. {Omotosho}, J.~S. {Mandeep}, A tropical
  model for analyzing radio refractivity: Selected locations in north central,
  nigeria, Proceeding of the 2015 International Conference on Space Science and
  Communication (IconSpace), 10-12 August 2015, Langkawi, Malaysia.

\bibitem{Ojo}
O.~L. Ojo, J.~S. {Ojo}, P.~{Akinyemi}, Characterization of secondary
  radioclimatic variables for microwave and millimeter wave link design in
  nigeria, Journal of Radio and Space Physics 46 (2017) 83--90.

\bibitem{Fashade}
O.~O. Fashade, T.~V. {Omotosho}, S.~A. {Akinwumi}, K.~P. {Olorunyomi},
  Refractivity gradient of the first 1km of the troposphere for some selected
  stations in six geo-political zones in nigeria, IOP Conf. Ser.: Mater. Sci.
  Eng. 640 (2019) 012087.
\newblock \href {http://dx.doi.org/10.1088/1757-899X/640/1/012087}
  {\path{doi:10.1088/1757-899X/640/1/012087}}.

\end{thebibliography}

\end{document}